\documentclass[journal,twoside]{IEEEtran}
\usepackage{cite}
\usepackage{amsmath,amssymb,amsfonts}
\usepackage{algorithmic}
\usepackage{graphicx}
\usepackage{textcomp}
\usepackage{wrapfig}

\def\BibTeX{{\rm B\kern-.05em{\sc i\kern-.025em b}\kern-.08em
    T\kern-.1667em\lower.7ex\hbox{E}\kern-.125emX}}
\def\journalname{IEEE Sensors Journal}

\begin{document}
\title{Uncooled Poisson Bolometer for High-Speed Event-Based Long-wave Thermal Imaging}
\author{Mohamed A. Mousa, Leif Bauer, Utkarsh Singh, Ziyi Yang, Angshuman Deka, and Zubin Jacob%
\thanks{Manuscript submitted Month XX, 2025. This work was partially supported by an Elmore Chaired Professorship at Purdue University.}
\thanks{M. A. Mousa, L. Bauer, U. Singh, Z. Yang, and Z. Jacob are with the Elmore Family School of Electrical and Computer Engineering, Purdue University, West Lafayette, IN 47907 USA (e-mail: zjacob@purdue.edu).}
\thanks{A. Deka is with the Birck Nanotechnology Center, Purdue University, West Lafayette, IN 47907 USA.}
\thanks{Corresponding author: Zubin Jacob (e-mail: zjacob@purdue.edu).}}

\maketitle

\begin{abstract}
Event-based vision provides high-speed, energy-efficient sensing for applications such as autonomous navigation and motion tracking. However, implementing this technology in the long-wave infrared remains a significant challenge. Traditional infrared sensors are hindered by slow thermal response times or the heavy power requirements of cryogenic cooling. Here, we introduce the first event-based infrared detector operating in a Poisson-counting regime. This is realized with a spintronic Poisson bolometer capable of broadband detection from 0.8-14$\mu\text{m}$. In this regime, infrared signals are detected through statistically resolvable changes in stochastic switching events. This approach enables room-temperature operation with high timing resolution. Our device achieves a maximum event rate of 1,250 Hz, surpassing the temporal resolution of conventional uncooled microbolometers by a factor of 4. Power consumption is kept low at 0.2$\mu$W per pixel. This work establishes an operating principle for infrared sensing and demonstrates a pathway toward high-speed, energy-efficient, event-driven thermal imaging.
\end{abstract}

\begin{IEEEkeywords}
Event-based vision, long-wave infrared (LWIR), spintronics, Poisson processes, uncooled infrared sensing, neuromorphic engineering, asynchronous readout, thermal detectors, low-latency imaging.
\end{IEEEkeywords}

\section{Introduction}
\label{sec:introduction}
\IEEEPARstart{L}{ong}-wave infrared (LWIR, 8–14 µm) sensing is essential for a wide range of fields, including wildfire detection \cite{sorenson2024thermal}, autonomous navigation \cite{bao2023heat}, and industrial surveillance \cite{rogalski2019two}. Most conventional LWIR focal plane arrays (FPAs) rely on a periodic, frame-based sensor (FBS) readout paradigm \cite{rogalski2022scaling}. This architecture can be inefficient because it captures redundant data in static scenes and limits the timing resolution to the fixed frame rate. These constraints often lead to high latency and excessive power consumption during high-speed tasks \cite{rogalski2022scaling}. Furthermore, while capturing every pixel in every frame generates significant data overhead, it does not necessarily improve the detection of fast-moving targets. Existing approaches, such as photonic preprocessing \cite{mousa2021toward} and machine-learning-driven computational imaging \cite{huang2024broadband}, have sought to mitigate these issues. Despite these advances, the underlying bottleneck of frame-based data acquisition remains a primary challenge for high-speed, energy-efficient thermal imaging.

Event-Based Vision Sensing (EVS) is a technology that can potentially address these limitations. Inspired by the human retina, EVS offers asynchronous, sparse, and low-latency readout \cite{gehrig2024low}. In an EVS, each pixel independently triggers an event when a local brightness change exceeds a predefined threshold. This produces a continuous stream of events rather than conventional frames \cite{gallego2020event}. Figure 1a illustrates the conceptual advantage of this paradigm. Conventional FBS are limited by discrete sampling in time, which reduces timing resolution and leads to motion blur in high-speed dynamic scenes. Additionally, these sensors produce large amounts of redundant data under static conditions. In contrast, an event-based system asynchronously reports only changes in scene brightness. This results in a sparse, high-speed data stream that captures sub-microsecond thermal dynamics.

Despite the advantages of event-based sensing, a significant spectral gap exists in the current technology landscape. As shown in the performance benchmarks in Figure 1b, current LWIR sensors generally fall into two categories. These include room-temperature microbolometers with limited response speed, and high-speed photon detectors (e.g., InSb or MCT) that require power-intensive cryogenic cooling \cite{razeghi2014advances}. Consequently, while event-based architectures are well-established for visible (Si) and short-wave infrared (SWIR) bands \cite{son20174, jakobson2022event}, the MWIR/LWIR regime lacks uncooled, low-power solutions (Figure 1c). To bridge this gap, we introduce the Spintronic Poisson Bolometer (SPB). By leveraging a different detection regime with an ultra-small active area ($0.05\ \mu\text{m}^2$), the SPB minimizes thermal mass and enables high-speed, room-temperature operation. This nanoscale footprint offers a pathway toward ultra-high-density focal plane arrays (FPAs) that surpass the scaling limits of conventional $12\text{--}17\ \mu\text{m}$ $VO_x$ microbolometers \cite{rogalski2022scaling}.

\begin{figure*}[htbp]
    \centering
    \includegraphics[width=0.9\textwidth]{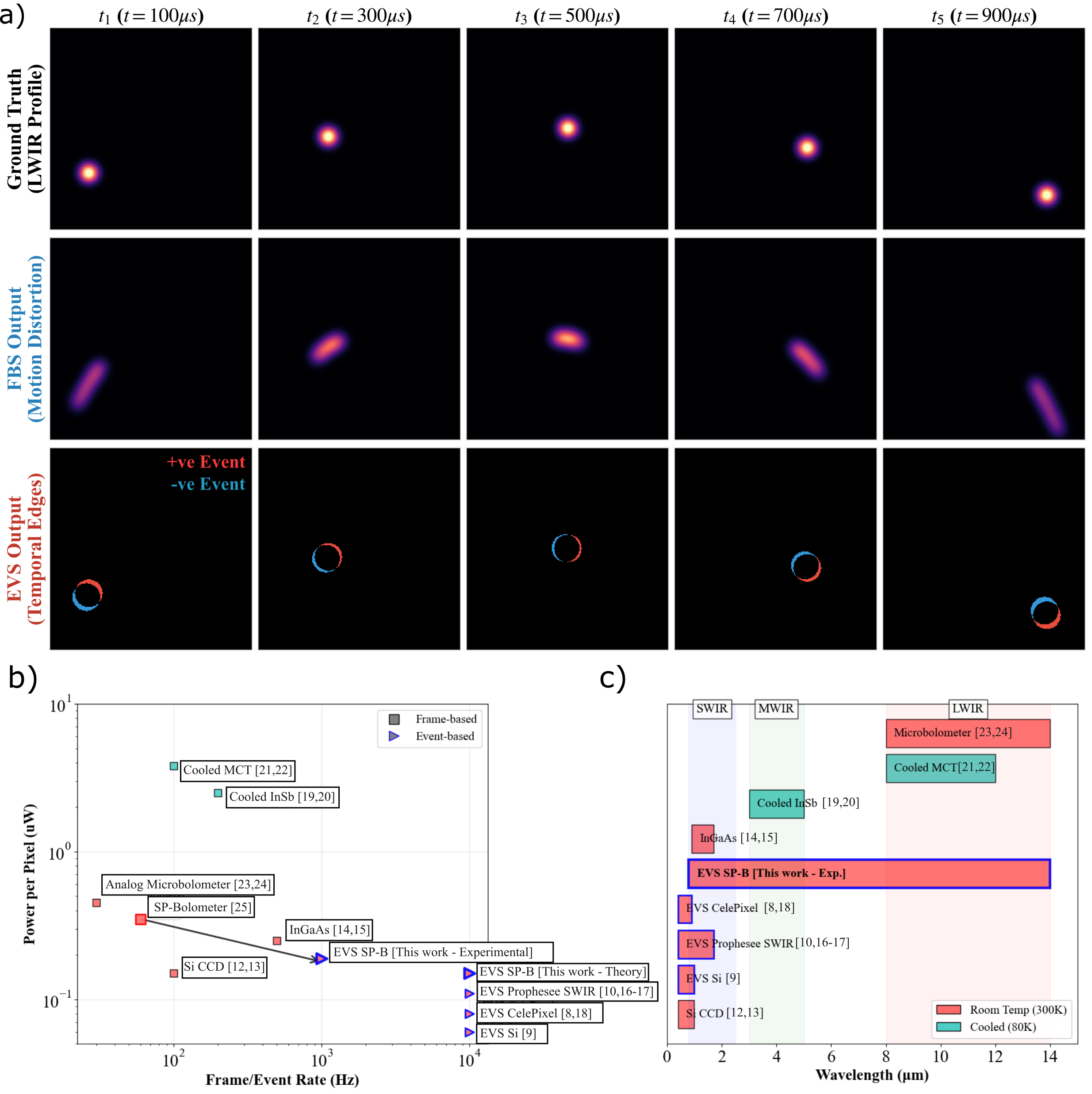}
    \caption{Conceptual Paradigm and Technology Landscape for LWIR Event-Based Sensing. (a) Comparison of a high-speed thermal target in a ballistic trajectory. The frame-based sensor (FBS) exhibits significant motion blur, whereas the event-based sensor (EVS) captures sharp, asynchronous temporal edges with sub-microsecond resolution. (b) Performance Benchmarking: Power consumption versus event/frame rate. EVS technologies are denoted by blue-outlined triangles and FBS by black-outlined squares. The spintronic Poisson (SP) bolometer (this work, thick blue outline) shows superior efficiency over conventional technologies. (c) Spectral Landscape: The SP-bolometer bridges the MWIR--LWIR gap. EVS solutions are identified by thick blue borders and FBS by black borders. Fill colors indicate operating temperature (red: 300\,K; teal: 80\,K). Data sources: Si CCD \cite{stevens2017recent,onsemi_KAI04050}, EVS Si \cite{son20174}, InGaAs \cite{geum2024highly,sensors_640CSX_2024}, Prophesee SWIR \cite{jakobson2022event,prophesee_packaged2024,sony_imx636}, CelePixel \cite{chen2019live,gallego2020event}, cooled InSb \cite{sato2024mid,scd_pelican_d_640}, cooled MCT \cite{lobre2025new,flir_x8580sls_datasheet}, analog microbolometer \cite{fusetto2023readout,lynred_pico640gen2_datasheet}, and SP bolometer \cite{bauer2025exploiting}.}
    \label{fig:Figure1}
\end{figure*}

\begin{figure*}[htbp]
    \centering
    \includegraphics[width=1\textwidth]{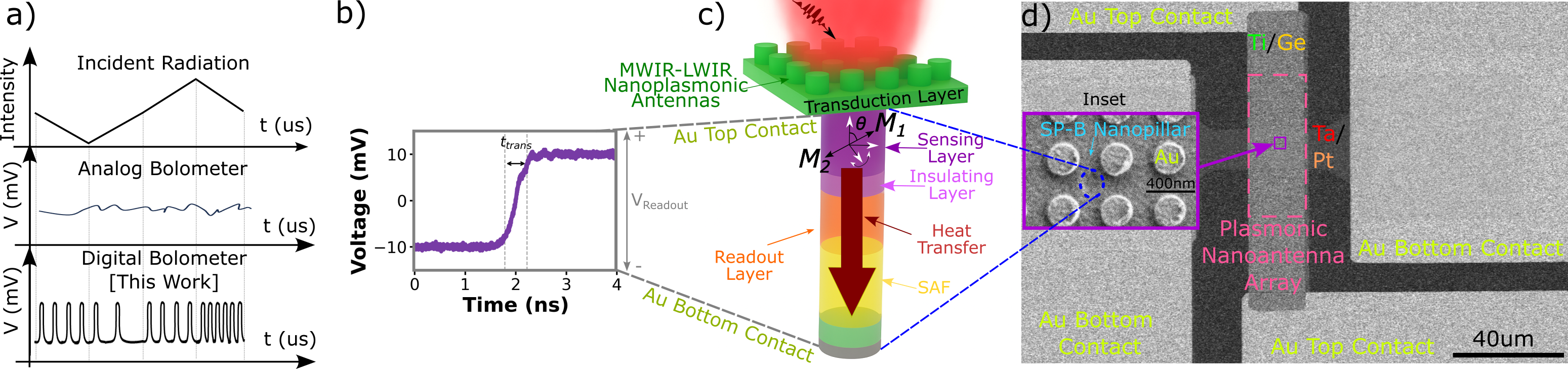}
    \caption{(a) Temporal response comparison between analog LWIR bolometers and the SP bolometer. The traditional analog response is effectively filtered by its high thermal inertia, whereas the SP bolometer resolves high-frequency dynamics by utilizing sub-nanosecond transition times. (b) Sub-nanosecond output transition demonstrating high-speed digital response. (c) Operating schematic of the SP bolometer at room temperature. Absorbed radiation generates a localized hotspot that propagates through the transduction layer, while a synthetic antiferromagnet (SAF) stabilizes the readout orientation. (d) SEM image of a single SP bolometer pixel. Inset: nanoplasmonic antenna.}
    \label{fig:Figure2}
\end{figure*}

This paper is organized as follows: Section \ref{sec:Concept} introduces the physical principles and architecture of the Spintronic Poisson Bolometer (SPB), detailing the stochastic transition mechanism. Section \ref{sec:experiment} provides an event-based characterization of the SPB, defining the differential count rate (DCR) and presenting a comparative performance analysis against conventional LWIR microbolometers. Section \ref{sec:Readout} describes the asynchronous pixel-level readout circuit design and presents spatiotemporal simulations of the system response to dynamic thermal scenes. A discussion of the data sparsity advantages and future multispectral integration is provided in Section \ref{sec:discussion}, followed by concluding remarks in Section \ref{sec:conclusion}.

\section{\label{sec:Concept}Spintronic Poisson Bolometer: Concept and Implementation}

The Spintronic Poisson Bolometer (SPB) operates on a sensing paradigm that redefines the transduction and readout of thermally absorbed energy. While the device remains governed by the physical constraints of heat absorption and thermal diffusion inherent to uncooled infrared detectors, it operates independently of the thermal time constant that limits conventional analog integration. Traditional microbolometers transduce temperature using continuous analog signals that require stabilization within a fixed readout frame. In contrast, the SPB architecture encodes thermal fluctuations into stochastic, thresholded switching events. By transitioning to this event-driven framework, the detection latency is decoupled from analog settling times and is instead defined by the statistical probability of a switching event.

The SPB moves away from traditional analog thermal sensing, where temperature is inferred from continuous resistance or voltage changes, towards a Poisson-counting regime \cite{bauer2025exploiting}.  Absorption of infrared radiation causes localized heating in the spintronic sensing layer, which increases thermal fluctuations that drive stochastic transitions between two stable magnetization states ($M_1$ and $M_2$ as shown in Figure 1c) \cite{hayakawa2021nanosecond,laughlin2019magnetic,coffey2012thermal}, producing discrete digital events whose rate encodes the thermal signal. At room temperature, these stochastic transitions arise from a carefully engineered small energy barrier ($E_b$) between the states, while an adjacent synthetic antiferromagnet (SAF) stabilizes the read-out layer’s orientation. The operating principle of the SPB is illustrated in Fig. 2c, where the thermal hotspot generated by absorbed infrared radiation propagates from the transduction layer through the device stack, increasing the probability of stochastic switching in the sensing layer \cite{kanai2021theory}.

Unlike individual photon detection, these events are driven by thermal fluctuations that are modeled as a Poisson process. In this framework, the probability of observing $N$ events within a fixed measurement interval $\tau$ is governed by a Poisson distribution. Under baseline room-temperature conditions, the mean number of counts is defined by $\langle N \rangle = \lambda_0 \tau$, where $\lambda_0$ represents the average event rate \cite{yang2025optical}. The probability of detecting $N$ stochastic events is expressed as:
\begin{equation} 
P(N; \lambda_0, \tau) = \frac{(\lambda_0 \tau)^N e^{-\lambda_0 \tau}}{N!}
\end{equation} 
Upon exposure to blackbody illumination of intensity $I_{\mathrm{BB}}$, the event rate increases from its baseline $\lambda_0$ to $\lambda_{\mathrm{BB}}$, which scales linearly with the absorbed infrared power. The resulting event statistics during the interval $\tau$ are described by:
\begin{equation}
P(N; \lambda_{\mathrm{BB}}, \tau) = \frac{(\lambda_{\mathrm{BB}} \tau)^N e^{-\lambda_{\mathrm{BB}} \tau}}{N!}
\end{equation}
In this framework, temperature is inferred from the shift in the average event rate rather than from continuous analog signal amplitudes. The temporal response of the SPB is therefore governed by the observation time required to statistically resolve a change in the rate $\lambda$, rather than by the constraints of thermal or electrical settling times.

The temporal performance and physical architecture of the SPB are summarized in Figure 2. A comparison of response times (Fig. 2a) highlights the device's ability to undergo sub-nanosecond digital transitions (Fig. 2b), a speed dictated by the ultrafast magnetization dynamics of the sensing layer. This contrasts sharply with the millisecond thermal time constants typical of standard microbolometers \cite{wang2025convergence}. 

In this framework, each pixel functions as a bistable threshold element where the threshold is defined by the physical energy barrier of the nanomagnet rather than an analog-to-digital comparison of signal amplitudes. These transitions are asynchronous, occurring at stochastic intervals governed by the instantaneous thermal state. While baseline thermal fluctuations manifest as a dark-event rate $\lambda_0$ (Eq. 1), infrared absorption increases this to a signal rate $\lambda_{\mathrm{BB}}$, forming the basis of the SPB’s EVS output. Because the mean event rate remains constant under static thermal flux, the architecture inherently eliminates redundant sampling, allowing the readout to prioritize pixels experiencing dynamic intensity shifts. 

The physical realization of this principle is shown in the SEM image of a SPB pixel (Fig. 2d), which features a nanoplasmonic antenna (inset) designed to enhance absorption across the 0.8--14 $\mu$m range. This integration of ultrafast physical switching with rate-based statistical sensing enables a truly frame-free, high-speed infrared imaging platform.

To validate the sensitivity of this event-driven approach, we measured the Noise Equivalent Temperature Difference (NETD). NETD is the key figure of merit for thermal imaging. It quantifies the minimum detectable temperature change. The experimental setup, utilizing a blackbody source and a ZnSe lens (f/0.5), is shown in Figure 3a. The SPB demonstrates a NETD of approximately 100 mK (Fig. 3b), proving that the move to a digital, stochastic readout does not compromise the high sensitivity required for thermal imaging applications. While the SPB demonstrates broadband physical absorption from $0.8$ to $14\ \mu\text{m}$ (as shown in Fig. 1c), we characterize the NETD specifically within the $3\text{--}14\ \mu\text{m}$ thermal atmospheric windows to evaluate its performance for mid- and long-wave infrared applications (Fig. 3c).

\begin{figure}[htbp]
    \centering
    \includegraphics[width=0.47\textwidth]{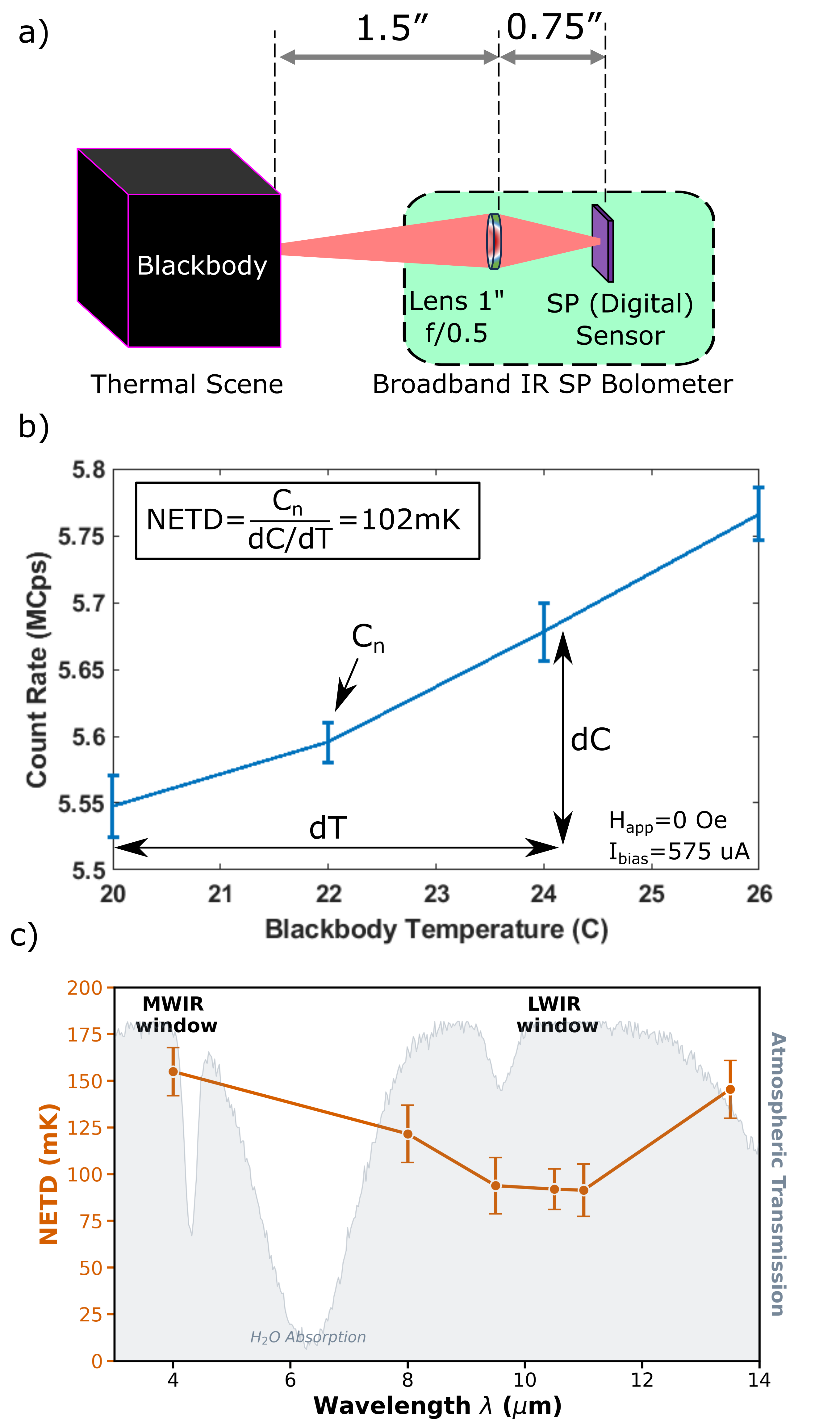}
    \caption{(a) Schematic of the NETD measurement setup for the SP-bolometer, including an ultra-stable blackbody source and a broadband ZnSe lens (f/0.5). (b) Measured count rate as a function of blackbody temperature, showing linear responsivity (dC/dT) used to calculate NETD = Cn/(dC/dT) = 102 mK, measured at $H_{app}$ = 0 Oe and $I_{bias}$ = 575 $\mu$A. (c) Spectral dependence of NETD across the mid-wave infrared (MWIR) and long-wave infrared (LWIR) atmospheric transmission windows, showing enhanced sensitivity in regions of high atmospheric transmission.}
    \label{fig:figure3}
\end{figure}

\begin{figure*}[htbp]
    \centering
    \includegraphics[width=0.9\textwidth]{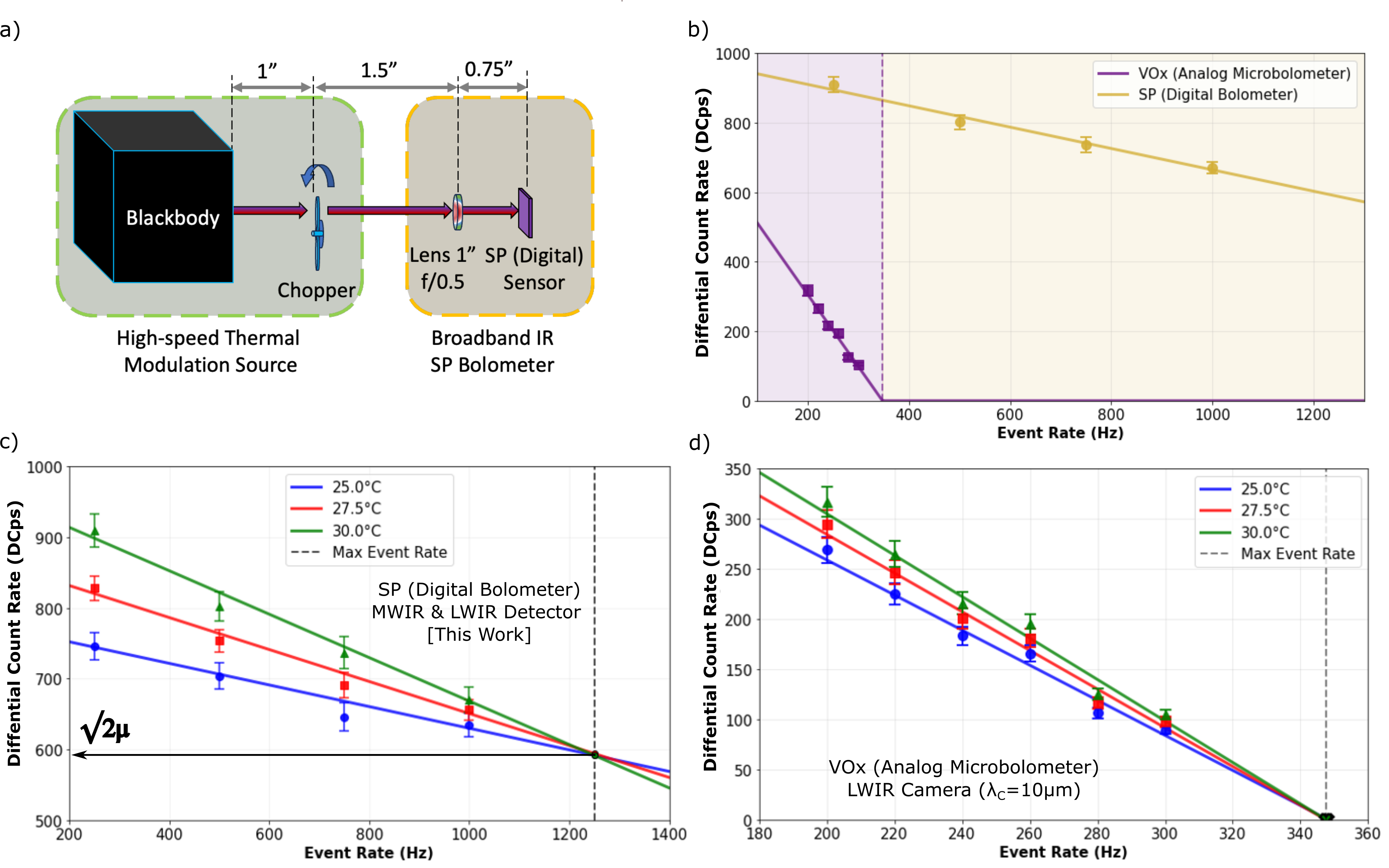}
    \caption{Event-based characterization of the SP bolometer. (a) Schematic of the EVS characterization setup, comprising an ultra-stable blackbody source ($<1$~mK stability), a mechanical chopper for high-speed thermal modulation, an f/0.5 ZnSe IR lens, the SP bolometer, and a bias magnet. (b) Differential count rate (DCR) comparison between the SP bolometer and a conventional sensor at a 30$^\circ$C blackbody temperature. The x-axis represents the chopper-driven modulation frequency, and the y-axis shows the differential count $D$ over a fixed integration window $\tau = 1$~ms. (c) DCR of the SP bolometer as a function of event rate for blackbody temperatures of 25, 27.5, and 30$^\circ$C, demonstrating reliable detection up to a maximum event rate of 1,250Hz. (d) Benchmarking DCR using a commercial uncooled LWIR camera (FLIR A325sc, featuring a $VO_x$ microbolometer FPA \cite{flirA325sc}) at 25, 27.5, and 30$^\circ$C. The FLIR sensor exhibits a maximum event rate of only 348Hz, constrained by its 8-12~ms thermal time constant and 60Hz internal readout electronics.}
    \label{fig:Figure4}
\end{figure*}

\section{\label{sec:experiment}Event-Based Characterization of the SP Bolometer}

To validate the high-speed temporal performance of the SP Bolometer, we characterized its response using the experimental configuration shown in Fig. \ref{fig:Figure4}a. Following established metrics for high-speed neuromorphic sensors \cite{gallego2020event, finateu20205}, we characterize the SPB using the Differential Count Rate (DCR). In our framework, DCR serves as the primary metric to bridge the gap between the underlying stochastic physics of the nanomagnet and the resulting asynchronous EVS output. Specifically, this metric quantifies the change in event frequency between consecutive integration windows of duration $\tau$, allowing us to map thermal fluctuations directly to digital event streams.

In this regime, infrared-induced thermal fluctuations are recorded as discrete detection events. Within a fixed window $\tau$, the total event count $N$ follows Poisson statistics. For a sensor with an intrinsic dark event rate $\lambda_0$ and a responsivity constant $\alpha$, the mean event count $\mu$ is approximately proportional to the incident blackbody flux $\Phi(T_{\mathrm{BB}})$:
\begin{equation}
\mu \approx (\alpha \Phi(T_{\mathrm{BB}}) + \lambda_0) \tau.
\end{equation}
The sensor output is generated as a differential count $D$ between two consecutive integration windows,
\begin{equation}
D = N_2 - N_1,
\end{equation}
where $N_1$ and $N_2$ are the counts measured in intervals $[t_0, t_0+\tau]$ and $[t_0+\tau, t_0+2\tau]$, respectively. This differential architecture serves as an intrinsic temporal filter. For a static scene where the flux remains constant, the expected value $\langle D \rangle = 0$, which effectively suppresses the static dark-count offset.

However, a change in the incident infrared power can occur between these two windows, such as that caused by a moving thermal object or a rapid temperature excursion. This change results in a shift from $\mu_1$ to $\mu_2$, yielding a non-zero mean $\langle D \rangle = \mu_2 - \mu_1$. Even in the static case, $D$ exhibits stochastic fluctuations arising from the underlying Poisson nature of the switching events. Since $D$ is the difference of two independent Poisson random variables, it follows a Skellam distribution with a variance defined by the sum of the means \cite{hwang2007}:
\begin{equation}
\mathrm{Var}(D) = \mu_1 + \mu_2.
\end{equation}

We refer to these stochastic variations in $D$ as \emph{differential fluctuations}, which define the shot-noise-limited floor of the DCR. In this architecture, the integration window $\tau$ determines the temporal resolution of the sensor. As $\tau$ is reduced to capture faster dynamics, the mean number of counts per window $\mu$ decreases, which consequently reduces the absolute magnitude of the differential signal $\langle D \rangle$ for a given flux change. However, because the SPB operates on discrete event statistics, the noise, represented by the standard deviation of the Skellam distribution, $\sigma_D = \sqrt{\mu_1 + \mu_2}$, scales down alongside the signal.

This behavior is distinct from conventional analog bolometers, where achieving high temporal resolution requires widening the measurement bandwidth, thereby increasing the integrated electronic Johnson and $1/f$ noise from the readout circuitry \cite{wang2024high}. While both analog and digital systems see an information reduction at shorter integration times, standard sensors suffer a secondary penalty from bandwidth-dependent electronic noise floors. In the SPB-based DCR, the noise is decoupled from the analog bandwidth of an amplification chain; instead, it remains strictly governed by the underlying Poisson event statistics. This allows the system to maintain a predictable signal-to-noise relationship ($SNR \propto \sqrt{\tau}$) even at high speeds, making it inherently suited for dynamic infrared sensing without the catastrophic SNR degradation often associated with wideband analog amplification.

The observed performance gap between the SPB and conventional microbolometers arises from two fundamentally different physical limits. For a standard $VO_x$ microbolometer, the response speed is intrinsically constrained by its thermal inertia, defined by the ratio of the pixel's heat capacity to its thermal conductance ($\tau_{th} = C/G$) \cite{huang2023ultra}. Consequently, flux modulation exceeding $1/\tau_{th}$ cannot be tracked by the pixel, which behaves as a first-order low-pass filter that suppresses high-frequency information \cite{wang2024high}. In contrast, as illustrated in Fig. \ref{fig:Figure4}c, the SPB DCR curves converge toward the Skellam noise floor. Experimentally, the standard deviation of these differential counts tracks the $\sqrt{\mu_1 + \mu_2}$ scaling, confirming that the readout is dominated by stochastic switching noise rather than electronic or thermal-lag components.

Crucially, while the pixel's thermal mass still dictates its ultimate responsivity, the sensing speed is not limited by a macroscopic thermal time constant. Instead, it is governed by the observation time required to achieve a target confidence interval for the rate $\lambda$. While the SPB still possesses a physical thermal mass, its ultra-miniaturized footprint ensures that the associated thermal settling time is orders of magnitude shorter than the temporal windows relevant to our measurements. Consequently, the SPB can be operated in a regime where performance is limited not by thermal inertia, but by the intrinsic statistical noise of the Poisson process. This enables reliable kilohertz-scale sensing at integration times well below the thermal limits of conventional microbolometers.

\section{\label{sec:Readout}Asynchronous Readout and Dynamic Scene Simulation}

\begin{figure}[htbp]
    \centering
    \includegraphics[width=0.48\textwidth]{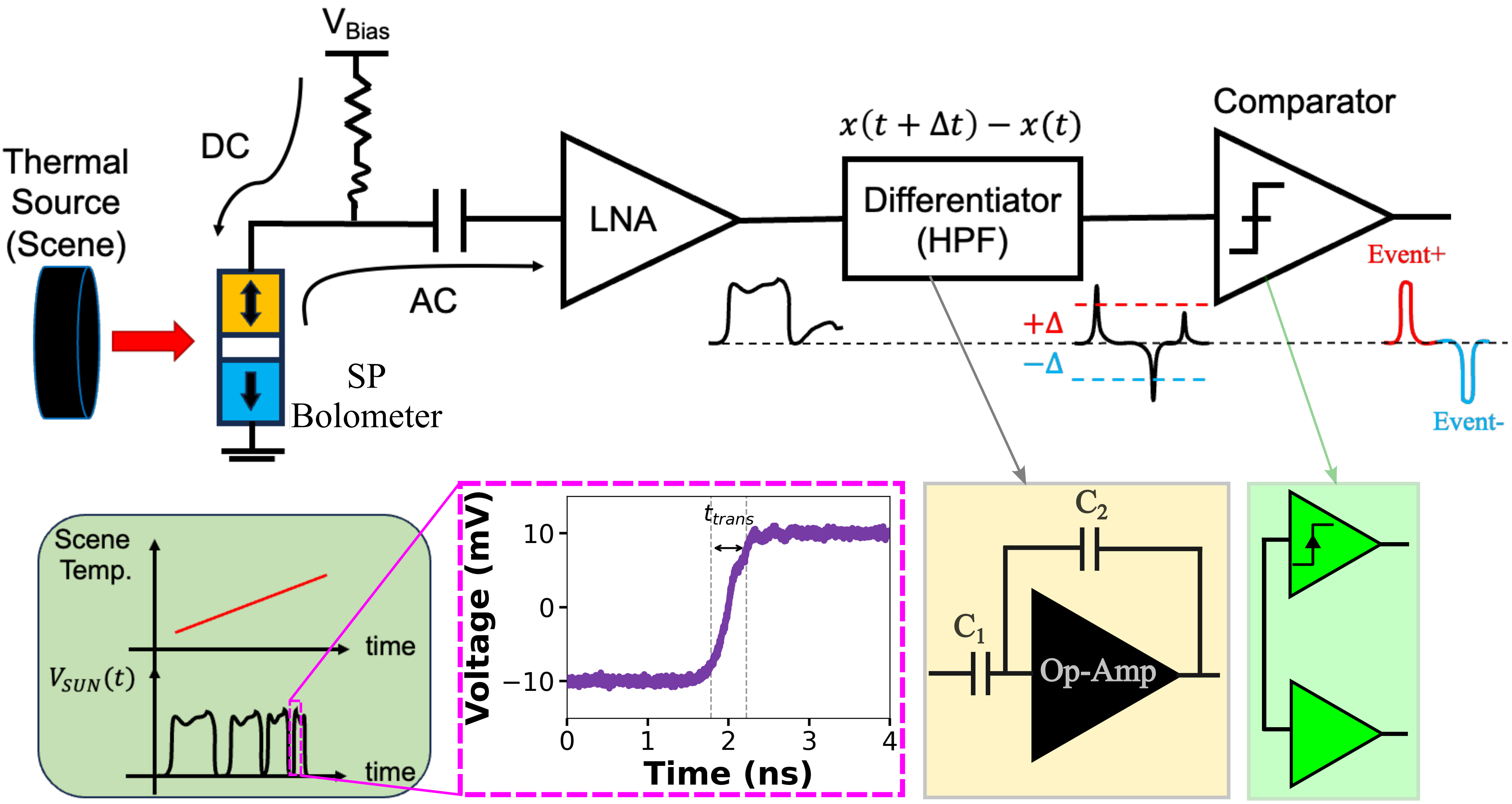}
    \caption{Circuit architecture of the SP bolometer event-based sensing (EVS) readout. The signal chain begins with the SP Bolometer, which converts incident thermal flux into a voltage signal. This signal is then AC-coupled to a Low-Noise Amplifier (LNA) to isolate the transient response. This is followed by a Differentiator (High-Pass Filter) that computes the temporal change in intensity, $x(t + \Delta t) - x(t)$, effectively suppressing static background. A Comparator stage then performs threshold detection against positive ($+\Delta$) and negative ($-\Delta$) limits to trigger asynchronous Event+ and Event- pulses. Insets illustrate the ultra-fast switching transition ($t_{trans} \approx 1$~ns) and the specific Op-Amp and comparator configurations used to achieve high-speed neuromorphic readout.}
    \label{fig:Figure5}
\end{figure}

\begin{figure}[htbp]
    \centering
    \includegraphics[width=0.48\textwidth]{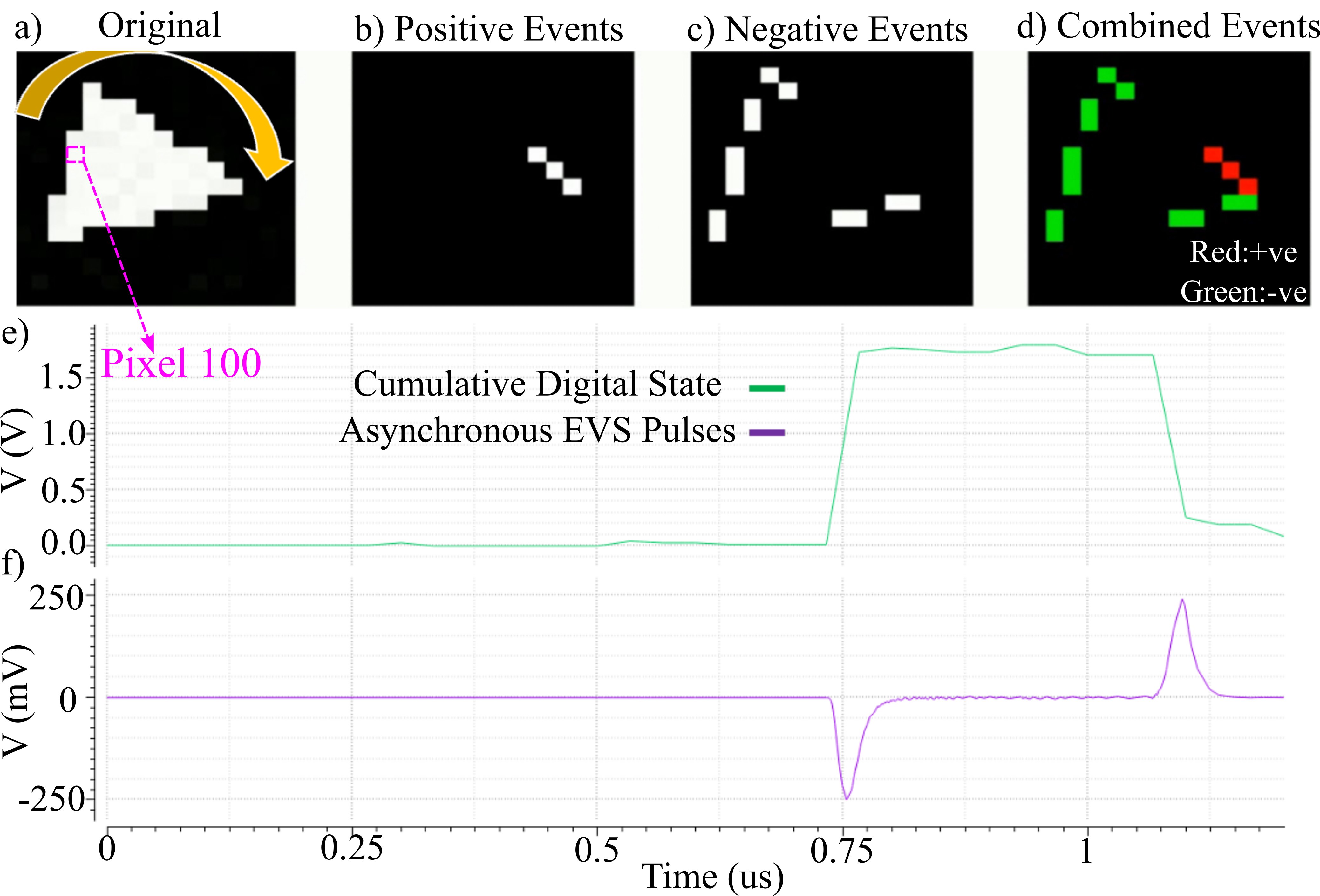}
    \caption{Circuit simulation of SP bolometer event-based sensing (EVS) readout. (a) Original 16$\times$16 synthetic triangular intensity pattern. (b) Detected positive (+ve), (c) negative (–ve), and (d) combined event maps generated by threshold crossings. (e) Time-domain digitized output for a representative pixel (Pixel 100). The green trace represents the cumulative digital state of the pixel. Transitions in this signal correspond to the accumulation of discrete stochastic events. (f) Corresponding asynchronous EVS readout pulses (purple trace) associated with individual threshold-crossing events. These spikes constitute the differential counts ($D$) generated at each magnetic switching transition, demonstrating that the SPB architecture intrinsically digitizes temporal flux changes into discrete, high-speed event packets.}
    \label{fig:Figure6}
\end{figure}

To translate these stochastic transitions into a usable EVS data stream, we developed a dedicated pixel-level asynchronous readout architecture. The circuit block diagram is shown in Fig. \ref{fig:Figure5}, comprising a low-noise amplifier (LNA), a high-pass differentiator for change detection, and a comparator that triggers time-stamped digital events. This architecture ensures that bandwidth is only consumed when the scene exhibits meaningful thermal dynamics. Detailed design parameters and circuit architecture are provided in the Supplementary Materials. The transition from a single-pixel device to a functional vision system was verified through physically-grounded circuit simulations (Fig. \ref{fig:Figure6}). Figure \ref{fig:Figure6}e and \ref{fig:Figure6}f demonstrate the single-pixel readout logic, where stochastic magnetization flips are successfully converted into discrete event spikes. When scaled to a 16 $\times$ 16 pixel array, the system effectively reconstructs dynamic scenes with high temporal fidelity.

\begin{figure*}[htbp]
    \centering
    \includegraphics[width=0.8\textwidth]{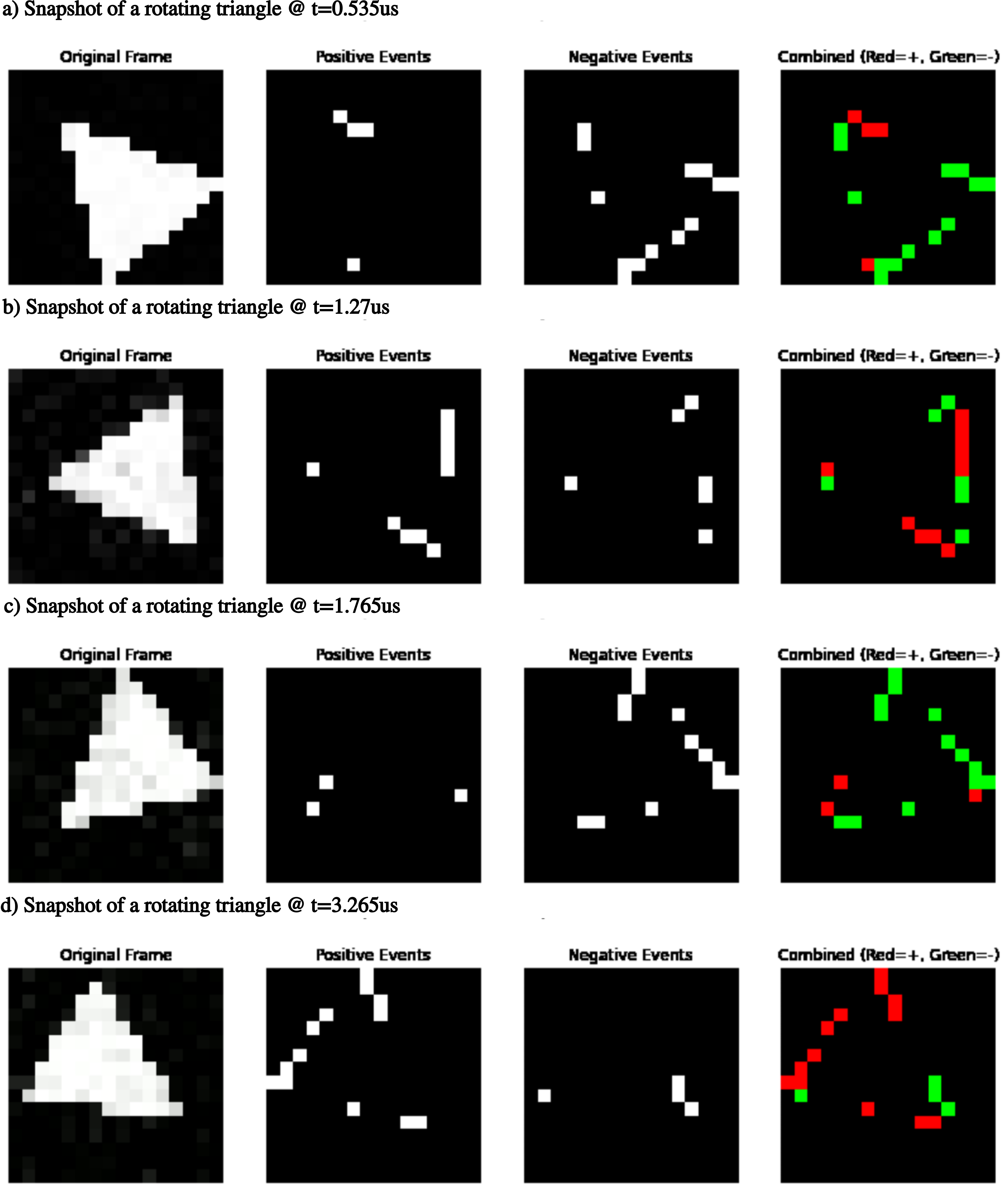}
    \caption{Event-based readout of a Spintronic Poisson bolometer. Circuit simulations show positive and negative Poisson events extracted from a 16x16 synthetic triangular scene at multiple time instances (red: +, green: -).}
    \label{fig:Figure7}
\end{figure*}

The spatiotemporal output for a synthetic moving scene is captured in Fig. \ref{fig:Figure6}a–d and further detailed in the sequence in Fig. \ref{fig:Figure7}. In these simulations, a triangular thermal target (Fig. \ref{fig:Figure6}a) moves across the field of view, triggering positive (ON) events at the leading edge (Fig. \ref{fig:Figure6}b) and negative (OFF) events at the trailing edge (Fig. \ref{fig:Figure6}c). The resulting combined event map (Fig. \ref{fig:Figure6}d) demonstrates the system’s ability to capture complex geometry and motion. Single-pixel digitized outputs (Fig. \ref{fig:Figure6}e) illustrate the accumulation of stochastic events over an integration window, while the corresponding analog readout pulses (Fig. \ref{fig:Figure6}f) show individual threshold-crossing events. These results confirm that the SPB-based event-driven sensing system resolves motion in the LWIR spectrum with sub-microsecond latency, a performance unattainable by conventional frame-based microbolometers. By providing asynchronous, digital-ready output at the pixel level, the SPB eliminates the need for power-intensive global shutters and high-bandwidth ADCs, enabling low-power, high-speed thermal machine vision.

\section{\label{sec:discussion}Discussion}

The realization of the Spintronic Poisson Bolometer (SPB) marks a departure from the traditional trade-offs in infrared thermography. The SPB reduces detection latency by replacing analog thermal settling and frame-based integration with a Poisson-counting readout. Although heat absorption and diffusion remain unchanged, the temporal resolution is set by statistical detectability rather than by thermal or electrical settling. While conventional sensors are often limited by the physical cooling rate of their pixel's thermal mass, the SPB’s asynchronous nature allows it to resolve transient thermal signatures with a temporal resolution exceeding 1.2kHz, a four-fold improvement over state-of-the-art commercial LWIR FPAs.

The primary advantage of this architecture is its data sparsity. In typical autonomous navigation or surveillance scenarios, the majority of the thermal background is static. Conventional frame-based sensors (FBS) consume power and bandwidth to digitize the redundant background repeatedly. In contrast, the SPB pixels remain silent until a statistically significant thermal change occurs, naturally compressing the data at the source. This makes the SPB an ideal front-end for neuromorphic engineering and edge-computing architectures, where low-power, event-driven inputs can be processed by Spiking Neural Networks (SNNs) without the latency of frame-to-tensor conversion \cite{mousa2025neural, cimarelli2025hardware}.

Future scaling of the SPB platform could leverage bio-inspired sensing strategies, such as micro-saccadic motion, to enhance spatial resolution in static scenes while maintaining high-speed temporal sensitivity \cite{gallego2020event}. Furthermore, the integration of broadband LWIR plasmonic absorbers or microlens arrays has demonstrated pathways for SPBs to achieve room-temperature sensitivities rivaling cooled detectors, significantly expanding the scope for high-performance thermal imaging \cite{singh2025long, yang2025design,yang2025long,mousa2026ultra}. Ultimately, the detector-independent nature of the Poisson-counting principle suggests this readout architecture could be hybridized with emerging 2D materials or quantum-dot absorbers to create multispectral EVS arrays spanning from the SWIR to the THz \cite{feng2024advances}.

\section{\label{sec:conclusion}Conclusion}

This work presented the Spintronic Poisson Bolometer (SPB), the first uncooled event-based sensor designed specifically for the MWIR/LWIR spectrum. By leveraging stochastic, thermally activated magnetization transitions, we have demonstrated a sensing modality that operates in a Poisson-counting regime, achieving broadband detection (0.8–14 µm) and event rates exceeding 1,250 Hz. Our experimental characterization and circuit simulations confirm that the SPB effectively decouples temporal resolution from analog thermal settling by encoding infrared-induced temperature fluctuations into stochastic digital events. Thus, it provides a low-latency, energy-efficient alternative to conventional frame-based detectors. Future work will focus on large-scale array integration, leveraging the spintronic nature of the device, which allows for seamless CMOS back-end-of-line (BEOL) compatibility. These results establish the SPB as a foundational technology for next-generation autonomous systems, high-speed biomedical diagnostics, and persistent environmental monitoring in power-constrained environments.

\section*{Acknowledgment}

This work was partially supported by an Elmore Chaired Professorship at Purdue University.

\bibliographystyle{IEEEtran}

\bibliography{jsen}

\end{document}